# The Orbital Period and Negative Superhumps of the Nova-Like Cataclysmic Variable V378 Pegasi

F. A. Ringwald*, Kenia Velasco, Jonathan J. Roveto, Michelle E. Meyers

Department of Physics
California State University, Fresno
2345 E. San Ramon Ave., M/S MH37
Fresno, CA 93740-8031

ABSTRACT

A radial velocity study is presented of the cataclysmic variable V378 Pegasi (PG 2337+300). It is found to have an orbital period of 0.13858 ± 0.00004 d (3.32592 ± 0.00096 hours). Its spectrum and long-term light curve suggest that V378 Peg is a nova-like variable, with no outbursts. We use the approximate distance and position in the Galaxy of V378 Peg to estimate $E(B-V) = 0.095$, and use near-infrared magnitudes to calculate a distance of 680 ± 90 pc and $M_V = 4.68 ± 0.70$, consistent with V378 Peg being a nova-like. Time-resolved photometry taken between 2001 and 2009 reveals a period of 0.1346 ± 0.0004 d (3.23 ± 0.01 hours). We identify this photometric variability to be negative superhumps, from a precessing, tilted accretion disk. Our repeated measurements of the photometric period of V378 Peg are consistent with this period having been stable between 2001 and 2009, with its negative superhumps showing coherence over as many as hundreds or even thousands of cycles.

KEYWORDS

cataclysmic variables – accretion disks – waves – AAVSO

* Corresponding author. Tel: +1-559-278-8426; fax: +1-559-278-7741.
*E-mail address*: ringwald@csufresno.edu

1. INTRODUCTION

Many properties of cataclysmic variable binary stars (CVs) depend on their orbital periods ($P_{orb}$), including their secular evolution, luminosity, and outbursts (Shafter et al., 1986). The catalogues of Patterson (1984) and of Ritter and Kolb (2003) include only CVs of known or suspected period, which emphasizes this parameter's importance.

Radial velocity studies reveal orbital periods more reliably than do photometric modulations, save for eclipses, but are not without problems. The strongest features in CV spectra are their emission lines, which originate in the CVs' accretion disks. Orbital

*1*

semi-amplitudes, or *K*–velocities, when measured from the emission lines, do not reliably trace the motion of the white dwarf, since they originate in the disk. Neither do $\gamma_{em}$, the emission line mean velocity, nor $T_0$, the epoch of spectroscopic phase zero. Eclipsing systems sometimes show $T_0$ lagging the eclipse by over 70 degrees in phase (e.g., Thorstensen et al., 1991). Reviews on CVs that discuss these issues include those by Warner (1995) and by Hellier (2001).

V378 Pegasi was discovered by the Palomar-Green survey (Green et al., 1986), and listed as PG 2337+300. The Palomar-Green survey classified V378 Peg as a hot subdwarf (sd). Koen and Orosz (1997) first recognized that V378 Peg is a cataclysmic variable, showing a light curve that flickers and a spectrum with a blue continuum and Balmer lines with emission cores flanked by absorption cores.

Wils (2011) discovered that V378 Peg has a sinusoidal light curve with an amplitude of 0.15 magnitudes and a period of either 0.1349 or 0.1560 days. In this paper we resolve the alias choice unambiguously, refine the period to 0.1346 ± 0.0004 days, and show that this variability is from negative superhumps (Harvey et al., 1994; see also Chapter 6.7 on page 92 of Hellier, 2001).

Table 1 lists the radial velocities we have measured. Table 2 presents the derived orbital parameters. In Section 2, observational procedure, data reduction, and analysis for the radial-velocity study are described. The spectrum is discussed in Section 3. The time-resolved photometry, our discovery of periodic variability, and our interpretation that this is from negative superhumps, are in Section 4. The absolute magnitude and distance of V378 Peg is estimated from near-infrared magnitudes in Section 5. The long-term light curve compiled by the AAVSO is discussed in Section 6, and our conclusions and summaries are in Section 7.

**Table 1.** Hα Emission Radial Velocities for V378 Peg, 2003 September 2-4[a]

| HJD[b] | V (km s$^{-1}$) | HJD[b] | V (km s$^{-1}$) | HJD[b] | V (km s$^{-1}$) |
|---|---|---|---|---|---|
| 2885.749 | 44 | 2885.958 | −30 | 2886.822 | 32 |
| 2885.758 | 15 | 2885.969 | −30 | 2886.829 | −5 |
| 2885.778 | −6 | 2885.976 | −20 | 2886.839 | −9 |
| 2885.786 | −10 | 2885.986 | −7 | 2886.847 | −15 |
| 2885.796 | −39 | 2885.994 | 2 | 2886.857 | 24 |
| 2885.804 | −37 | 2886.004 | −13 | 2886.864 | 8 |
| 2885.814 | −4 | 2886.012 | 20 | 2886.873 | 81 |
| 2885.821 | −25 | 2886.684 | −69 | 2886.881 | 8 |
| 2885.831 | −3 | 2886.692 | −39 | 2886.89 | −31 |
| 2885.839 | 3 | 2886.701 | −25 | 2886.898 | −13 |
| 2885.847 | −5 | 2886.709 | −6 | 2886.915 | 2 |
| 2885.854 | −9 | 2886.718 | 17 | 2886.923 | −28 |
| 2885.864 | 13 | 2886.726 | 34 | 2886.932 | −27 |
| 2885.871 | 6 | 2886.735 | 10 | 2886.94 | −45 |
| 2885.881 | 28 | 2886.743 | 13 | 2886.949 | −25 |
| 2885.888 | 21 | 2886.753 | −16 | 2886.956 | −35 |
| 2885.898 | 17 | 2886.761 | 10 | 2886.966 | −32 |
| 2885.906 | 44 | 2886.77 | −9 | 2886.973 | 0 |
| 2885.916 | −20 | 2886.778 | −22 | 2886.984 | 17 |
| 2885.924 | −28 | 2886.788 | −1 | 2886.991 | 84 |
| 2885.933 | −5 | 2886.795 | 3 | 2887.001 | 70 |
| 2885.941 | 14 | 2886.805 | −39 | 2887.009 | −1 |
| 2885.951 | −22 | 2886.812 | −29 | | |

[a] Velocities were measured by fitting Gaussians to the Hα emission line profile.
[b] Barycentric Julian Date of mid-integration, minus 2,450,000.

**Table 2.** Derived Orbital Parameters for V378 Peg, from Hα velocities[a]

| $P_{orb}$ (days) | $K_{em}$ (km s$^{-1}$) | $\gamma_{em}$ (km s$^{-1}$) | $T_0$ (HJD − 2,450,000) | $\sigma$ (km s$^{-1}$) |
|---|---|---|---|---|
| 0.13858 ± 0.00004 | 23.9 ± 2.8 | -3.4 ± 2.0 | 2885.854 ± 0.004 | 23 |

[a] Velocities fitted to $V(t) = \gamma_{em} + K_{em} \sin[2\pi (t - T_0)/P_{orb}]$. All velocities were measured by fitting Gaussians to the Hα emission lines. All errors are estimated to 68% confidence (see Thorstensen and Freed, 1985).

## 2. RADIAL-VELOCITY OBSERVATIONS

The radial-velocity study was carried out on 2003 September 2-3 and 3-4 UT, with spectra taken with the 2.1-m telescope at Kitt Peak National Observatory and its Gold spectrograph and F3KB Ford (3072 x 512) CCD. A GG 495 order-sorting filter was used in the spectrograph, important since cataclysmic variables are so blue. The long-slit spectra permitted accurate sky subtraction through the 2.0-arcsecond slit, used for all exposures. Instrument rotation was not used.

The wavelength coverage spanned about $\lambda\lambda$ 403.0 – 897.0 nm at 0.6-nm resolution, at 0.247 nm per channel dispersion. Exposure times were all 600 seconds. Spectra of HeNeAr lamps were taken every 20 minutes, between pairs of spectra, throughout the observations. Flat-field exposures were taken with a quartz lamp inside the spectrograph.

All spectra were reduced and analyzed as in Thorstensen and Freed (1985) and Ringwald et al. (1994). Radial velocities were measured by fitting the H$\alpha$ emission line to a Gaussian, and taking the wavelength of the centroid of the Gaussian to correspond to the velocity of the line. A Lomb-Scargle periodogram (Lomb, 1976; Scargle, 1982; Press et al., 1992) of the radial velocities is shown in Figure 1. The correct alias is obvious, at a period of 0.13858 ± 0.00004 days (3.32592 ± 0.00096 hours).

Figure 2 shows a least-squares fit to a sinusoid with the most likely period. Spectrograph flexure caused variations on the order of ± 10 km s$^{-1}$, shown by velocities measured from the night-sky line at $\lambda$ 557.7 nm and plotted in Figure 2.

## 3. AVERAGE SPECTRUM

An average of the individual spectra, comprising 8.25 hours' observing time, is shown in Figure 3. These spectra have been normalized, or divided at each channel by the intensity of the continuum there, because we did not take flux standards, since the purpose of the observation was to measure radial velocities, through a narrow slit.

This average spectrum of V378 Peg resembles that of a dwarf nova in outburst, or of a nova-like variable (Williams, 1983; Figure 3.6 of Hellier, 2001), with relatively weak Balmer lines in emission on a bright blue continuum. He I $\lambda$ 587.6 nm and He I $\lambda$ 667.8 nm are in weak emission, and He I $\lambda$ 447.1 nm is in absorption.

Koen and Orosz (1997) show a similar spectrum, with absorption wings flanking cores, characteristic of non-magnetic nova-like variables (Williams 1983). There is no sign of absorption features such as TiO bands from the mass-losing secondary star, nor of He II $\lambda$ 468.6 nm emission, to indicate strong magnetism, although the CIII/NIII blend at $\lambda$ 464.0 nm is weakly present in emission.

## 4. TIME-RESOLVED PHOTOMETRY AND NEGATIVE SUPERHUMPS

*Observations*

Observations of V378 Peg were made on the nights of 2008 October 22-24, 2008 November 21-23, 2009 September 17-19, 2009 September 23-26, 2009 November 23-25 and 2009 November 28-30 from Fresno State's station at Sierra Remote Observatories, located at the Sierra Nevada Mountains of central California. The equipment used was a DFM Engineering 16-inch, f/8 telescope with an SBIG STL-11000M camera. All observations were taken through an Astrodon Clear luminance filter, which admits most of the visible range between 350nm and 1000 nm. Weather was clear and photometric for nearly all observations: we estimate that this site gets 200 photometric nights per year, with most of them between May and October.

The observations consisted of taking a series of images. Each image was exposed for 60 seconds, with 7 seconds of dead-time between exposures, to read out the CCD. All images were calibrated with darks only, but the photometry should be accurate to within 1%, which we estimate from the scatter in the *V−C1* plots (see Figure 4). This telescope and camera have an image scale of 0.51 arcseconds/pixel. All our observations were binned 3x3, making for an image scale of 1.53 arcseconds/pixel. The seeing ranged between 0.87 and 2.15 arcseconds on all nights.

Additional time-resolved differential photometry of V378 Peg was taken on the nights of 2001 August 1-3 UT in service mode with the 40-inch telescope at Mount Laguna Observatory, its CCD 2001 camera, and no filter. This telescope and camera have an image scale of 0.40 arcseconds/pixel, and all exposures were binned 4x4, making for an image scale of 1.62 arcseconds/pixel. The weather was clear, and seeing was between 1.9 and 2.3 arcseconds.

*Photometric measurements*

All magnitudes were measured with the Astronomical Image Processing for Windows (AIP4WIN) version 2 software (Berry and Burnell 2005). We used differential aperture photometry, to measure a variable star and a star of constant brightness, called the comparison star, denoted by C1. Then by comparing their values, tiny changes in the brightness of the variable can be detected. In addition a second comparison star, denoted by C2, of similar brightness to the variable, is measured and the magnitude difference of the two comparison stars is computed to create a second light curve. This second light curve is used as a check. For all observations, the star used as C1 was GSC 2766:1864, which has $V = 13.67$. Also for all observations, the star used as C2 was GSC 2766:1304, with $V = 14.15$. V378 Peg itself is also listed in the *Hubble Space Telescope Guide Star Catalog* (Lasker et al., 1990), as GSC 2766:1346 with $V = 13.72$.

All photometry we measured used an aperture centered on the star by the software with a diameter of 6 pixels, or 3.06 arcseconds. The software also counted photoelectrons in an annulus centered on the star by the software between 9 and 12 pixels (4.59 and 6.12 arcseconds) in diameter, for sky subtraction.

*Light Curves and Lomb-Scargle Periodograms*

Once the photometric data was obtained it was used to create light curves and Lomb-Scargle periodograms. A light curve is a graph of the variation of brightness of a celestial object as a function of time. A Lomb-Scargle periodogram is a discrete Fourier transform that is modified to find sinusoids in unevenly spaced data, which is the case for most astronomical datasets. It is basically a graph of power (in arbitrary units) as a function of frequency and it is used to find periodicities of celestial objects.

Figure 4 shows the light curve for all nights we observed V379 Peg in 2009 September. This figure shows the magnitude difference between V378 Peg (denoted by *V*) and comparison star *C1* and the magnitude difference between the two comparison stars, *C1* and *C2*. Figure 4 shows that comparison star C1 certainly has a constant magnitude, whereas the magnitude of V378 Peg is varying. Figure 4 also shows that V378 Peg has a saw-tooth shaped light curve, which shows periodic variability. These characteristics are present in light curves from observations dating back to 2008 and 2001, which are shown in the Appendix (available electronically only).

The graph in Figure 5 shows the Lomb-Scargle periodogram of our photometry of V378 Peg during the seven nights we observed it during 2009 September. The strongest peak in the power spectrum is at a frequency of $7.419 \pm 0.017$ cycles/day, which corresponds to a period of $3.2350 \pm 0.0074$ h. The peaks on either side of this peak are one-cycle-per-day aliases—false peaks that are produced due to the gaps in the light curve because of daylight (see pages 130-131 of Heller 2001).

Our radial velocity study above measured the orbital period of V378 Peg at $3.32592 \pm 0.00096$ h. The photometric period found in the 2009 September observations is 2.73% shorter than the orbital period. Periods that are found to be a few percent different from the orbital period are superhumps. We therefore interpret the photometric modulation to be negative superhumps, which implies that V378 Peg is a novalike, since it shows no obvious dwarf nova outbursts (see Section 6 below and the AAVSO light curve shown at the end of the Appendix, available electronically), and its spectrum (see Figure 3 and Koen and Orosz, 1997) shows a strong blue continuum with weak emission lines.

*Coherence of the Negative Superhump Period*

We also checked the consistency of the negative superhump period. The periods for each observation set is shown in Table 3. The negative superhump period did not vary significantly from 2001 to 2009.

**Table 3.** The negative superhump period of V378 Peg

| Observation Set | Frequency (cycles/day) | Period (hours) |
|---|---|---|
| 2001 August | 7.4187 ± 0.0011 | 3.2351 ± 0.0002 |
| 2008 October | 7.423 ± 0.067 | 3.233 ± 0.029 |
| 2008 November | 7.441 ± 0.069 | 3.225 ± 0.030 |
| 2009 September | 7.419 ± 0.017 | 3.2350 ± 0.0074 |
| 2009 November | 7.407 ± 0.025 | 3.242 ± 0.011 |

With observations taken during five epochs, we can study the coherence of the photometric period by calculating the quality factor $Q = \Delta f/f$. Table 3 shows that the periods measured during the different observations of the negative superhumps were consistent with no change in period at all, over as much as eight years. Table 4 compiles calculations of $Q$ from our measured periods, although we stress that these should be viewed as upper limits, since we did not observe V378 Peg continuously.

**Table 4.** Coherence in the negative superhump frequencies of V378 Peg

| Observation Sets | $Q = f/\Delta f$ |
|---|---|
| 2001 August to 2009 November | 634 |
| 2001 August to 2008 October | 1730 |
| 2008 October to 2008 November | 4130 |
| 2008 November to 2009 September | 340 |
| 2009 September to 2009 November | 6170 |

*Precessional Period*

The nodal precession period of the disk was calculated using the equation (Hellier 2001),

$$\frac{1}{P_{nsh}} = \frac{1}{P_{orb}} + \frac{1}{P_{nodal}}$$

where $P_{nsh}$ is the negative superhump period, $P_{orb}$ is the orbital period and $P_{nodal}$ is the nodal precession period. The above equation yielded $P_{nodal}$ = 4.96 days = 119 hours. There is excess noise at low frequencies in our periodograms (see Figure 5), although there isn't sufficient frequency resolution to state definitely whether we have detected this precessional period. More time-resolved photometry, over more nights, more contiguously, will be needed to do this.

5. NEAR-INFRARED MAGNITUDES, ABSOLUTE MAGNITUDE, AND DISTANCE

V378 Peg has $J = 13.775 \pm 0.032$, $H = 13.687 \pm 0.040$, and $K_s = 13.647 \pm 0.050$, in the 2MASS photometric system (Cutri et al., 2003). These near-infrared magnitudes and the method of Ak et al. (2007) implies a distance $d = 680 \pm 90$ pc for V378 Peg. This assumes that $E(B-V) = 0.095$ for V378 Peg, which we estimate using the method of Ak et al. (2008) and the interstellar dust maps of Schlegel et al. (1998), and also assuming that $A_J = 0.887\, E(B-V)$ (Fiorucci and Munari, 2003).

This distance implies an absolute magnitude $M_J = 4.54 \pm 0.70$, which is consistent with V378 Peg being a nova-like variable. With an average $V = 13.91$ from the AAVSO light curve (shown at the end of the Appendix), this implies that $M_V = 4.68 \pm 0.70$. This is consistent with the relation between absolute magnitude at maximum and orbital period of Warner (1987), which predicts $M_V = 4.78$, and of Harrison et al. (2004), which predicts $M_V = 4.65$.

6. LONG-TERM LIGHT CURVE

We have obtained data from the AAVSO database (Henden 2009) to produce the long term light curve (shown at the end of the Appendix). This light curve contains observational data from 1998 January 06 to 2009 October 25. We have plotted only definite detections: we have excluded observational limits from non-detections. We notice that there are no outbursts present in the long term light curve, thus indicating that V378 Peg is a nova-like. We cannot, however, rule out stunted outbursts discovered in some nova-likes by Honeycutt et al. (1998) and Honeycutt (2001), particularly near MJDs 2000 and 3200.

7. CONCLUSIONS

This paper presents a definitive orbital period of $3.32592 \pm 0.00096$ hours for V378 Peg, measured with a radial velocity study. The average spectrum shows a bright, blue continuum with relatively weak Balmer emission lines.

V378 Peg shows photometric variability we interpret to be negative superhumps, with a period of $3.23 \pm 0.01$ hours. This is 2.71% shorter than the orbital period, therefore it must be the negative superhump period. Besides exhibiting a powerful peak with a frequency of 7.42 cycles/day, the Lomb-Scargle periodograms. The nodal precession period was calculated to be 119 hours. Some power at corresponding low frequencies is present, but a definitive measurement of the nodal precession period will need a longer photometric dataset.

The distance of V378 Peg was calculated from near-infrared magnitudes by two methods. The method of Ak et al. (2007) gives a distance of 680 ± 90 pc and $M_V$ = 4.68 ± 0.70. The method of Beuermann (2006) gave a distance of 510 ± 70 pc and $M_V$ = 5.69 ± 0.30. We also estimate $E(B-V)$ = 0.095, from the approximate distance and position in the Galaxy of V378 Peg.

We have determined that V378 Peg is a nova-like cataclysmic variable. The reasons to suppose this are: (1) an optically thick spectrum characteristic of a CV in a high state, (2) negative superhump periods occur only in nova-likes, (3) the absolute magnitude corresponds to values typical of nova-likes, and (4) the long term light curve, compiled by the AAVSO, showing no outbursts with amplitudes greater than about one magnitude.

ACKNOWLEDGMENTS

This research used photometry taken at Fresno State's station at Sierra Remote Observatories. We thank Dr. Greg Morgan, Dr. Melvin Helm, Dr. Keith Quattrocchi, and the other SRO observers for creating this fine facility.

We thank Paul Etzel and Lee Clark for carrying out the 2001 photometric observations at Mount Laguna Observatory, which is operated by the Department of Astronomy at San Diego State University.

The spectra were taken at Kitt Peak National Observatory, which is part of the National Optical Astronomy Observatories, which are operated by the Association of Universities for Research in Astronomy, Inc., under cooperative agreement with the National Science Foundation. We thank Dave Reynolds and John Prigge, for help during the spectroscopic observations. The spectra were reduced with IRAF, the Image Reduction and Analysis Facility, which is distributed by the National Optical Astronomy Observatories.

This publication makes use of data products from the Two Micron All Sky Survey, which was a joint project of the University of Massachusetts and the Infrared Processing and Analysis Center/California Institute of Technology, funded by the National Aeronautics and Space Administration and the National Science Foundation. This research has made use of the NASA/IPAC Extragalactic Database (NED), which is operated by the Jet Propulsion Laboratory, California Institute of Technology, under contract with the National Aeronautics and Space Administration. This research has made use of NASA's Astrophysics Data System.

This research has made use of the Simbad database, which is maintained by the Centre de Données astronomiques de Strasbourg, France. We acknowledge with thanks the variable star observations from the AAVSO International Database contributed by observers worldwide and used in this research.

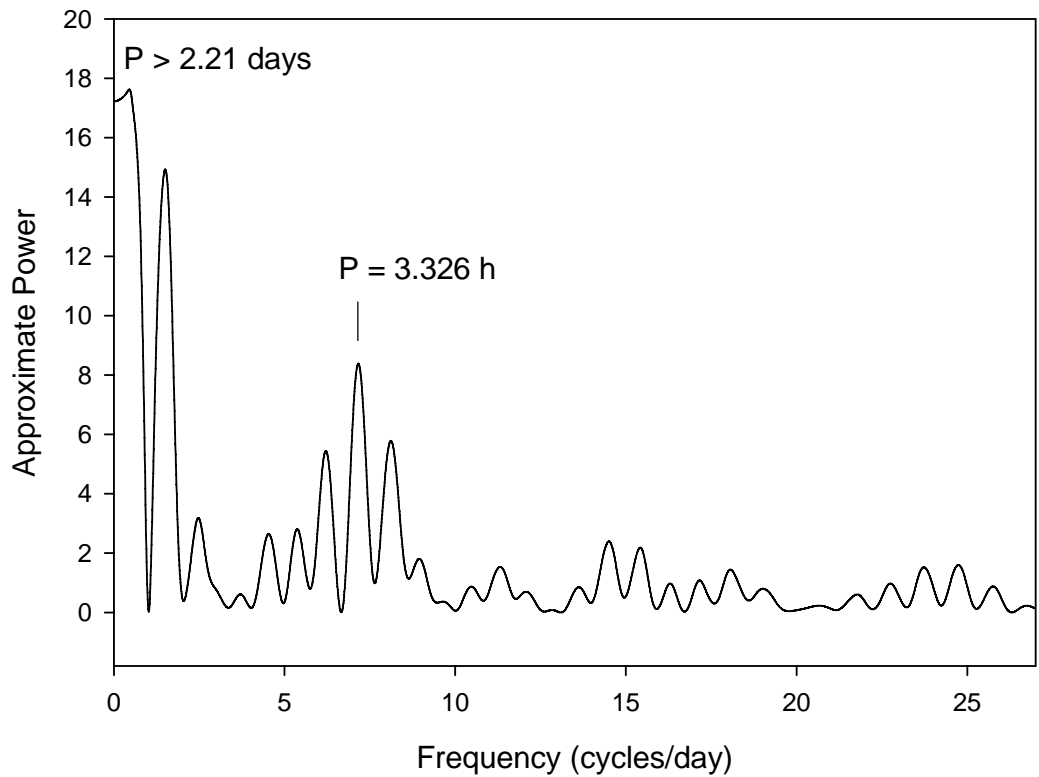

Figure 1. The Lomb-Scargle periodogram for the Hα velocities for V378 Peg. The highest peak shows the likely orbital period to be 0.1386 days (3.326 hours). Unresolved low-frequency noise is also labeled.

*12*

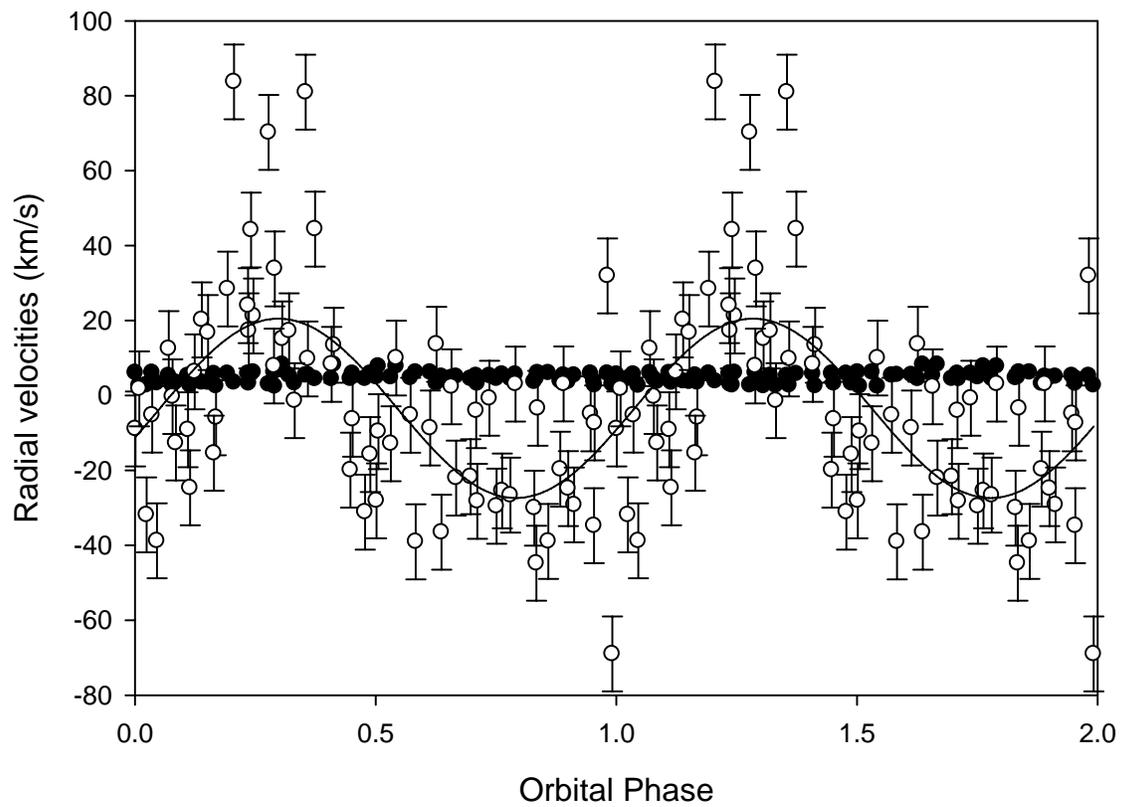

Figure 2. The radial velocity curve of the Hα velocities of V378 Peg. All velocities are plotted twice for continuity, using open circles, with error bars estimated by the Gaussian fits to the line profiles. Also plotted with filled circles are velocities measured from the night-sky line at $\lambda$ 557.7 nm, illustrating the amount of spectrograph flexure.

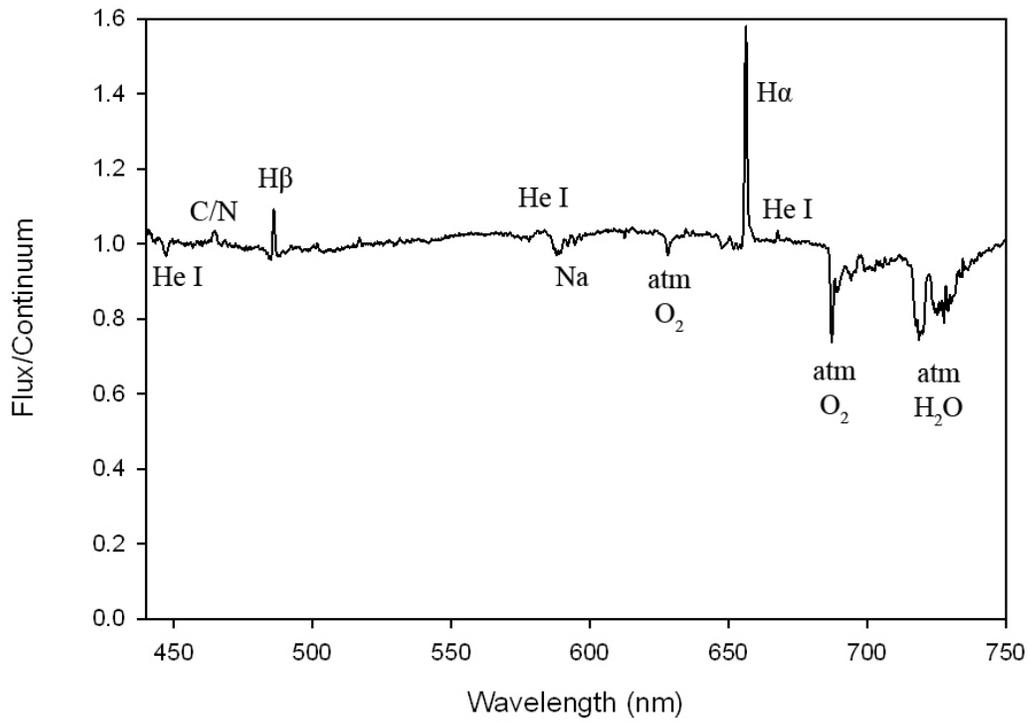

Figure 3. The average of the spectra of V378 Peg, comprising 8.25 hours' exposure time. The absorption features labeled "atm" are from Earth's atmosphere: all other labeled features are from V378 Peg.

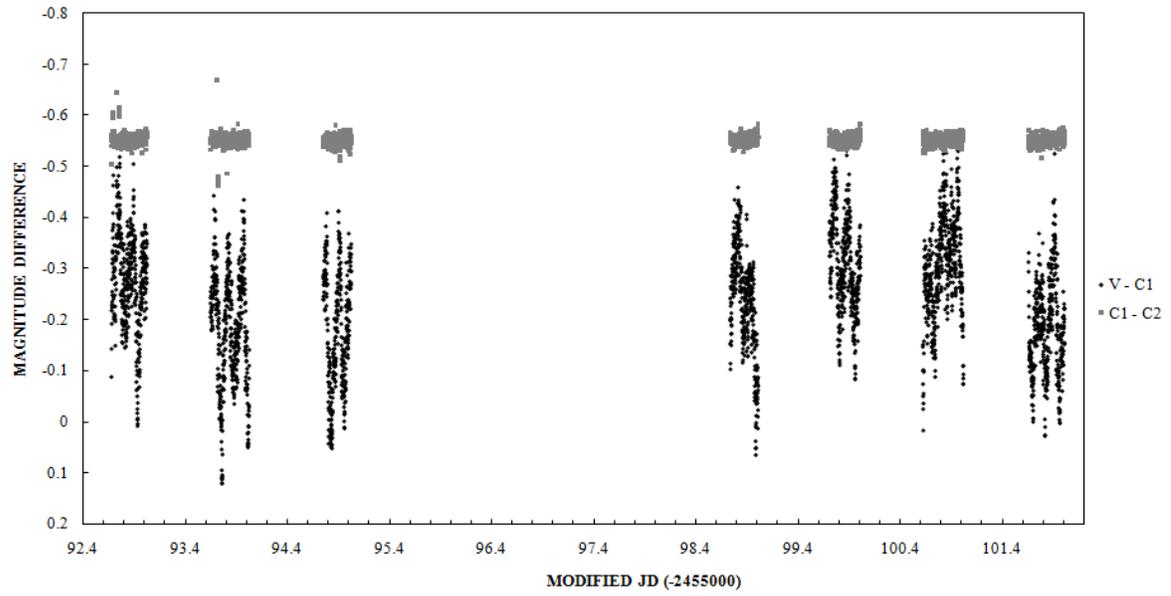

Figure 4. The light curves for all the observational nights of 2009 September of the magnitude difference between V378 Peg (denoted as V) and comparison star C1 and of the magnitude difference of the two comparison stars, C1 and C2.

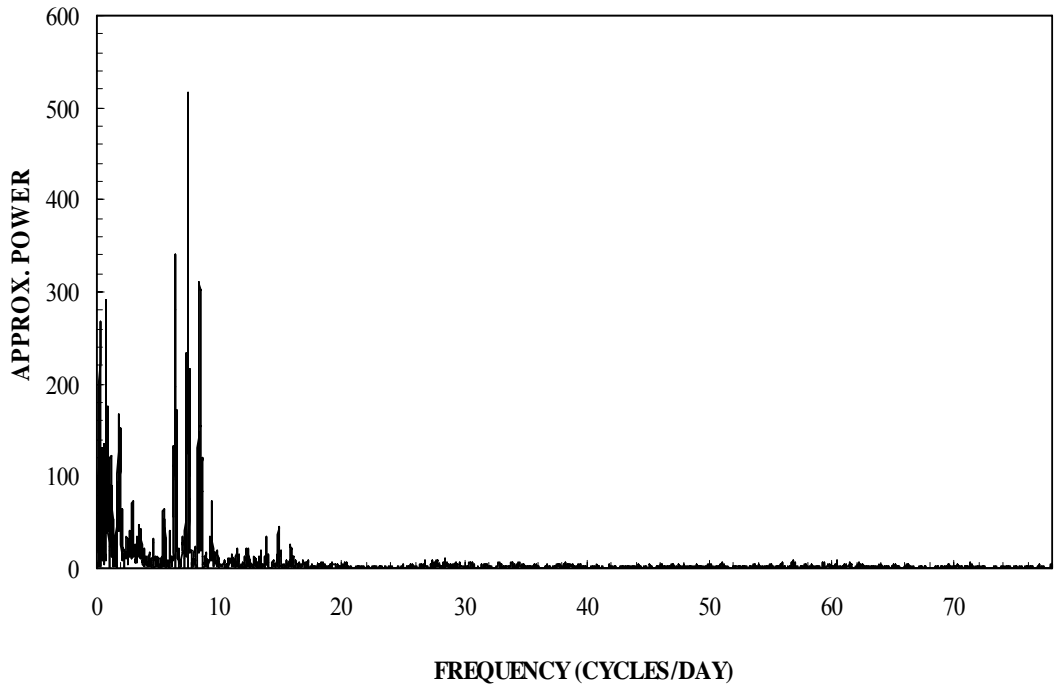

Figure 5. The Lomb-Scargle periodogram of the photometric data from 2009 September 17-19 and 2009 September 23-26 of V378 Peg

**APPENDIX:** supplemental figures from *The Orbital Period and Negative Superhumps of the Nova-Like Cataclysmic Variable V378 Pegasi,* by Ringwald, Velasco, Roveto, and Meyers.

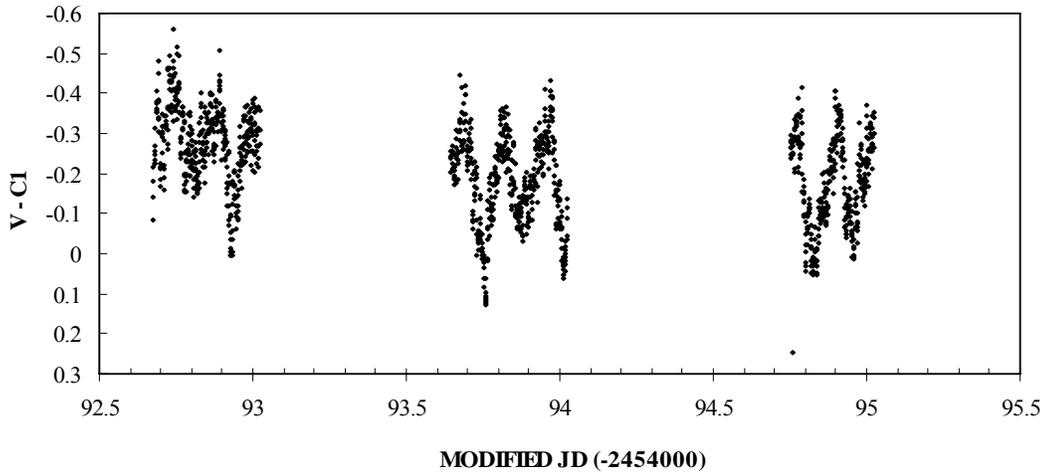

Figure 6. The light curve of the 2009 September 17-19 observations of V378 Peg

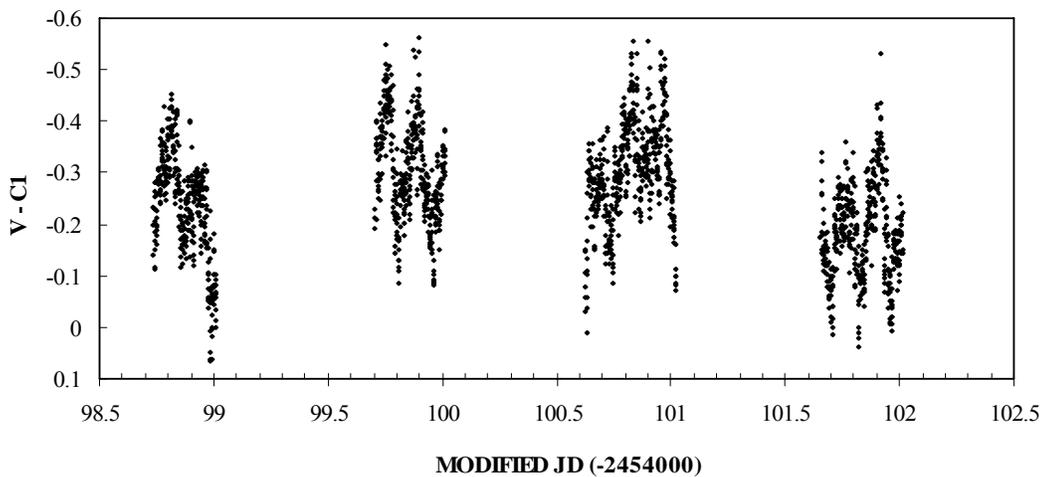

Figure 7. The light curve of the 2009 September 23-26 observations of V378 Peg

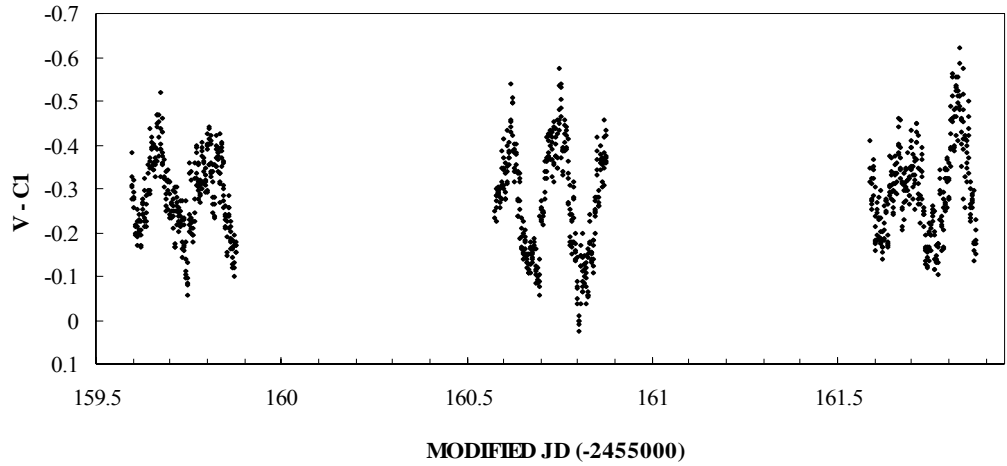

Figure 8. The light curve of the 2009 November 23-25 observations of V378 Peg

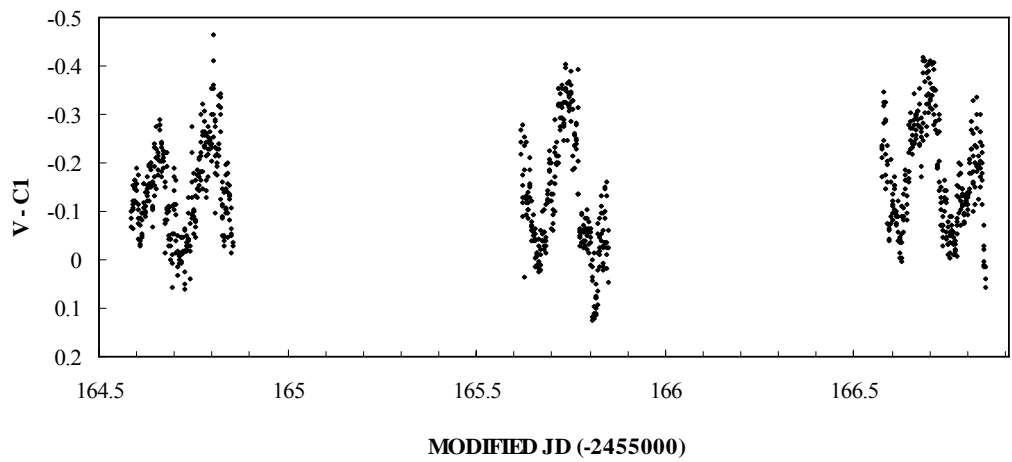

Figure 9. The light curve of the 2009 November 28-30 observations of V378 Peg

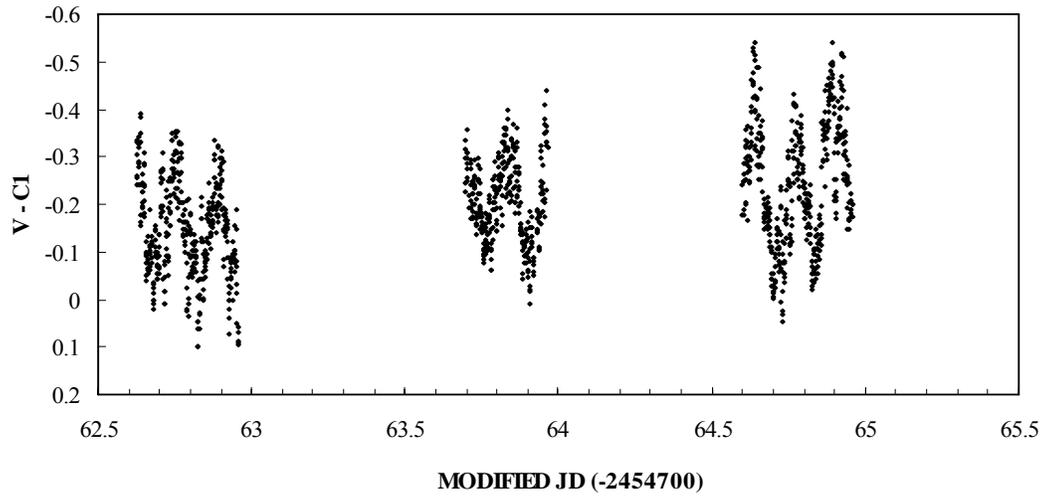

Figure 10. The light curve of the 2008 October 22-24 observations of V378 Peg

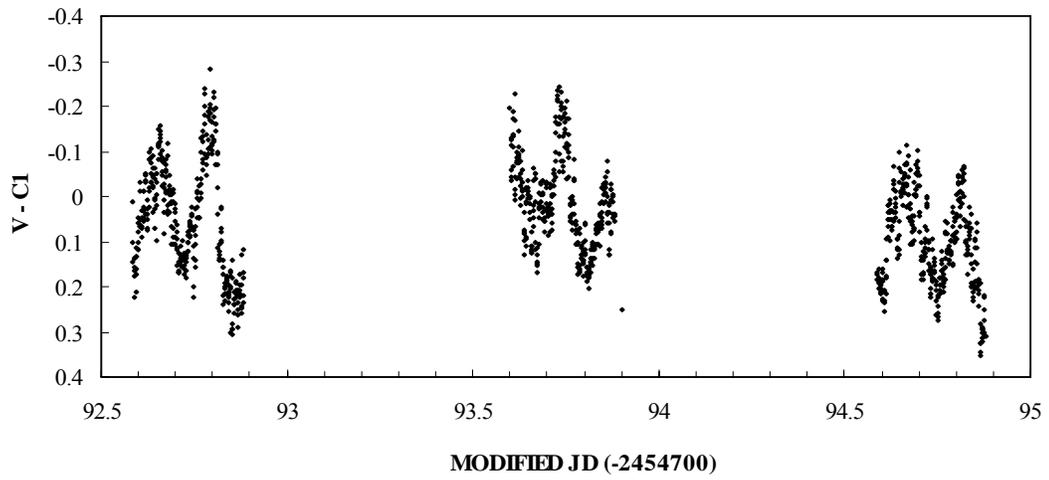

Figure 11. The light curve of the 2008 November 21-23 observations of V378 Peg

*19*

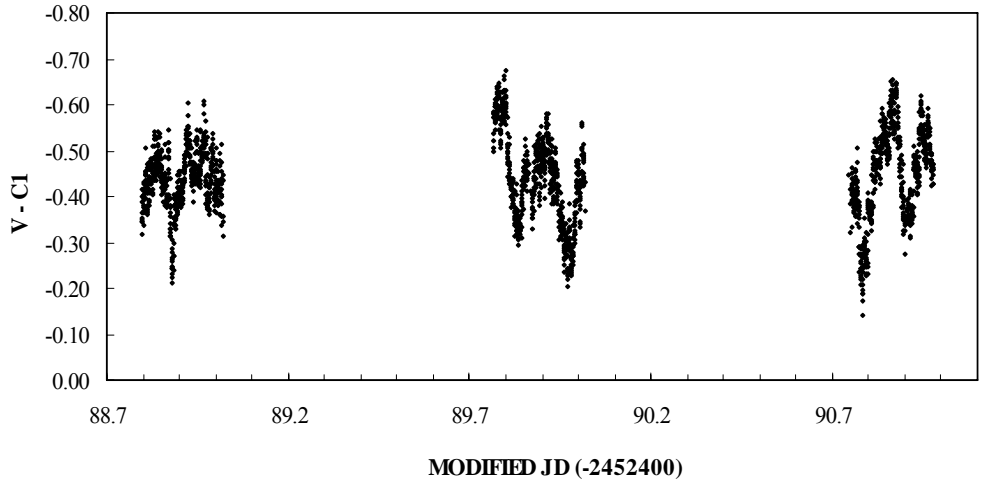

Figure 12. The light curve of the 2001 August 1-3 observations of V378 Peg

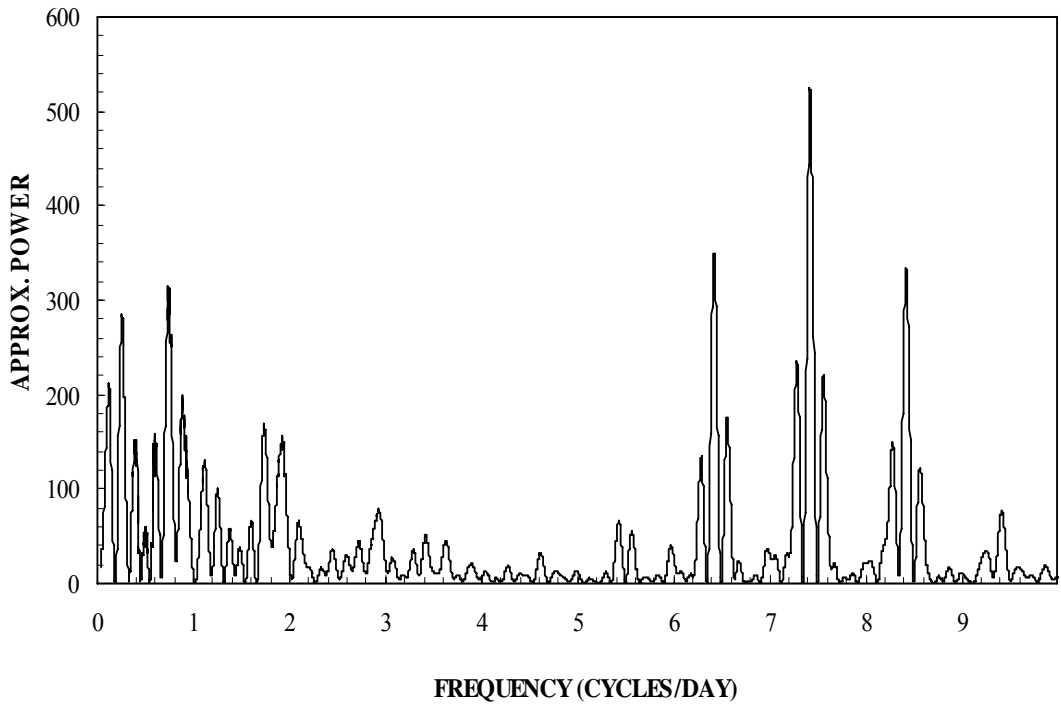

Figure 13. A closer look of Figure 5.

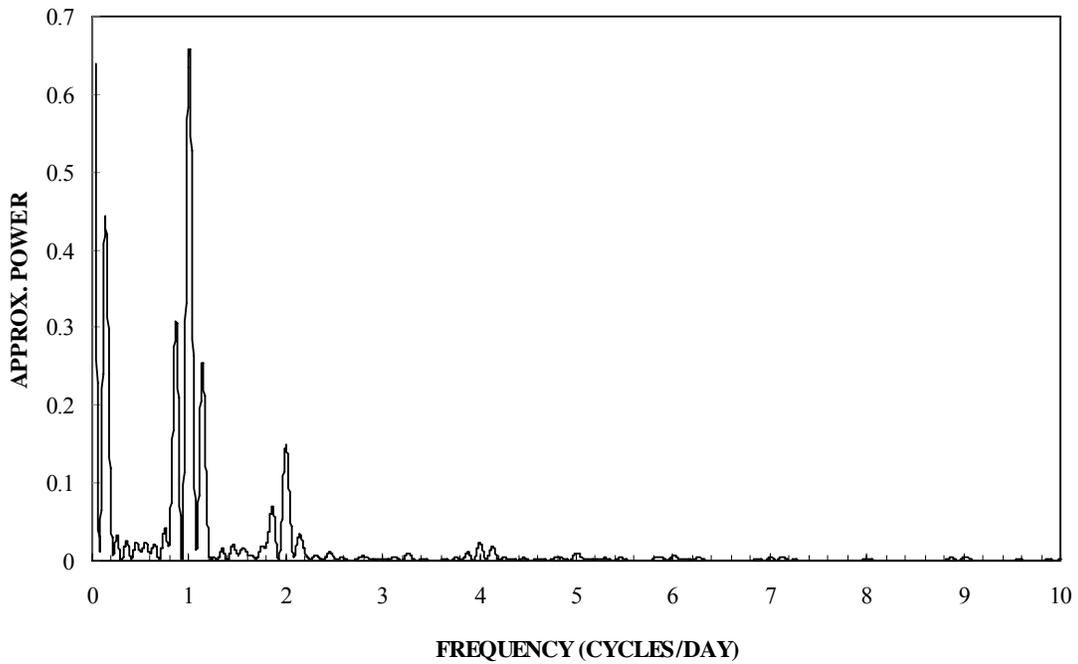

Figure 14. The window function of the photometric observations of 2009 September for V378 Peg.

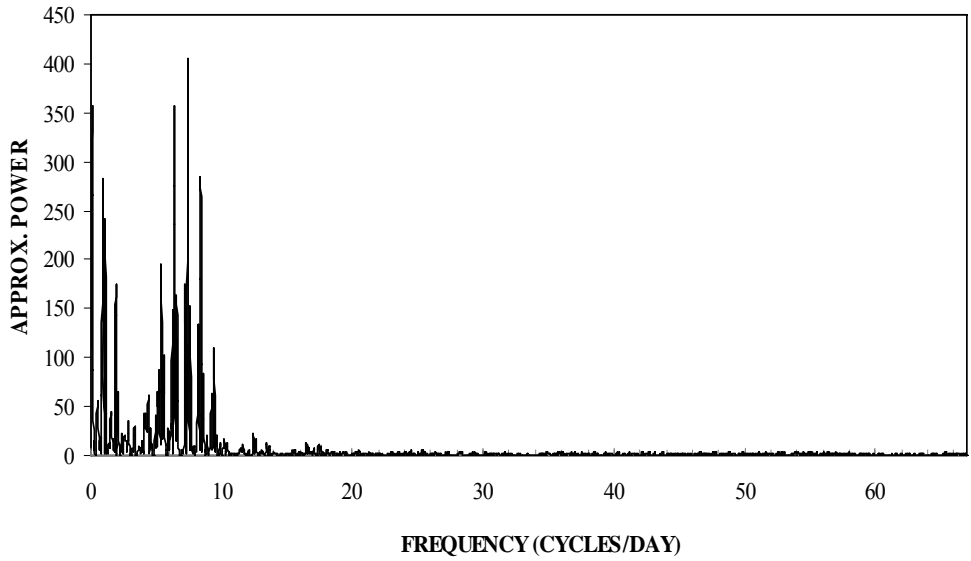

Figure 15. The Lomb-Scargle periodogram for the 2009 November 23-25 and 2009 November 28-30 observations of V378 Peg

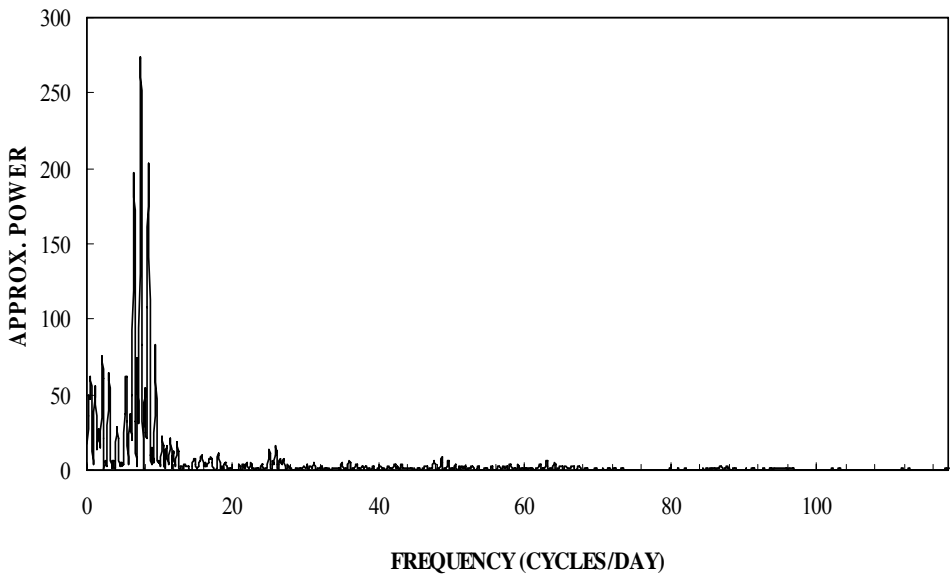

Figure 16. The Lomb-Scargle periodogram for the 2008 November 21-23 observations of V378 Peg

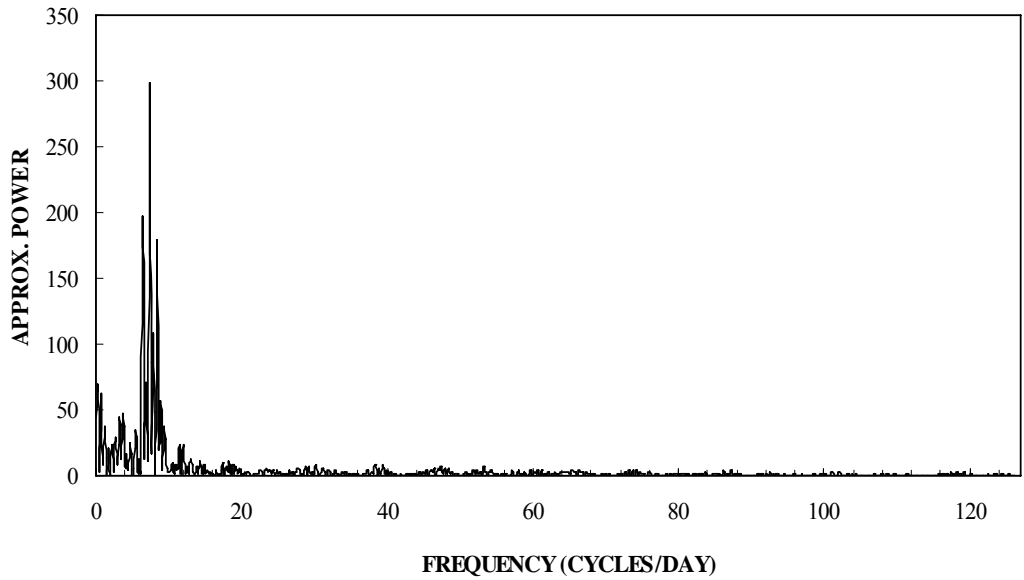

Figure 17. The Lomb-Scargle periodogram of the 2008 October 22-24 observations of V378 Peg

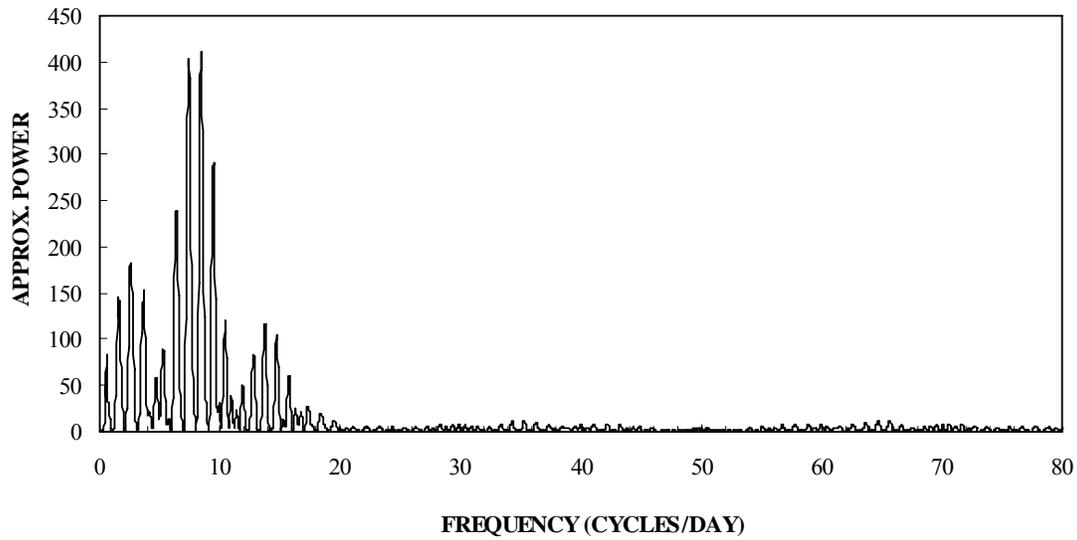

Figure 18. The Lomb-Scargle periodogram of the light curve of the 2001 August 1-3 of V378 Peg shows ambiguity, with approximately equal power at frequencies of 7.41871 cycles/day and 8.42749 cycles/day. We take the 7.41871 cycle/day alias to be true period, since it is consistent with the photometric period in 2008 and 2009. It is unsurprising that this time series suffered aliasing problems, since the longest series of observations in any one night in 2001 was 5.6 hours long, much shorter than the 8- and 9-hour/night series taken in 2008 and 2009.

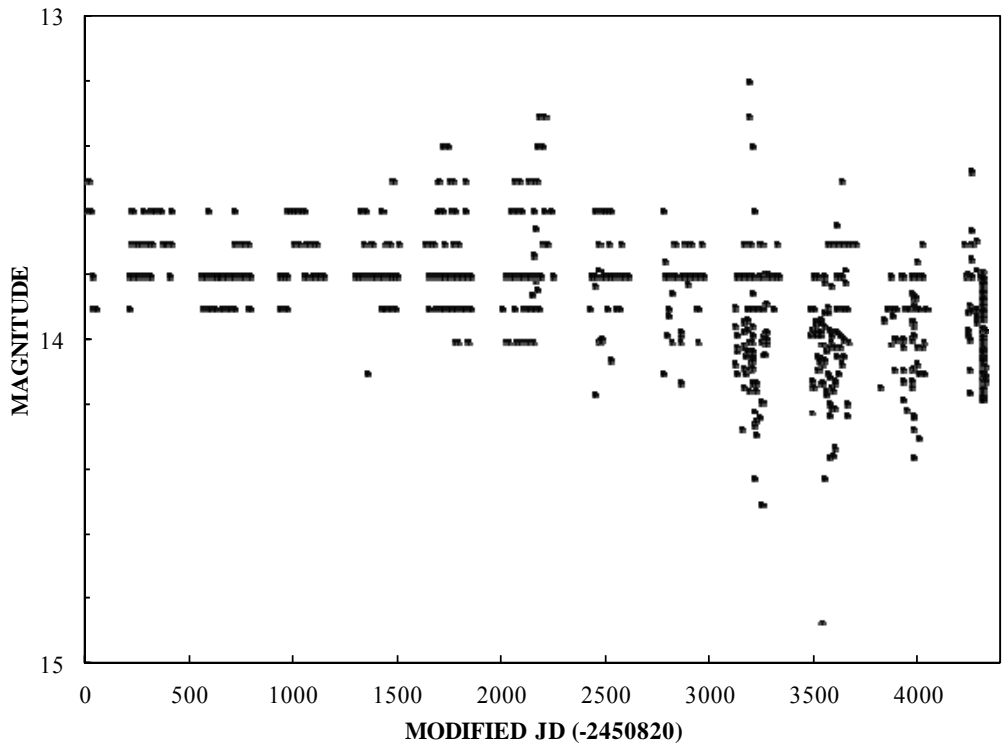

Figure 19. Long term light curve of V378 Peg, from 1998 January 6 to 2009 October 25, compiled by the AAVSO (Henden 2009).